\documentclass[10pt]{article}
\newcommand{\mc}{\multicolumn}

\usepackage{amsmath,amsfonts}
\usepackage{amsbsy}
\usepackage{color,soul}
\usepackage{verbatim}
\usepackage{setspace}
\usepackage{rotating}
\usepackage{array}
\usepackage{graphicx}
\usepackage{pdflscape}
\usepackage{booktabs}
\usepackage{colortbl}
\usepackage{xcolor}
\usepackage{multirow}
\definecolor{lightgray}{gray}{0.9}
\usepackage{xcolor}

\usepackage{enumerate}

\def\mR{\mathbb{R}}

\def\tx{{\tilde x}}

\def\definedas{\stackrel{\Delta}{=}}

\newcommand{\cX}{{\mathcal X}}

\newcommand{\cN}{{\mathcal N}}

\newcommand{\cL}{{\mathcal L}}

\newcommand{\cZ}{{\mathcal Z}}

\def\tZ{{\tilde Z}}

\def\real{\mR}
\def\bc{\begin{center}}
\def\ec{\end{center}}
\definecolor{lightgray}{gray}{0.9}

\newcommand{\beq}{\begin{eqnarray}}
\newcommand{\eeq}{\end{eqnarray}}
\newcommand{\beqq}{\begin{eqnarray*}}
\newcommand{\eeqq}{\end{eqnarray*}}

\title{Comparison of Persistence Diagrams}
\author{Sarit Agami\\
Department of Economics\\
Hebrew University, Mount Scopus, Jerusalem, Israel\\
email:sarit.agami@mail.huji.ac.il}

\begin{document}
\maketitle


%

\begin{abstract}
Topological Data Analysis (TDA) is an approach to handle with big data by studying its shape. A main tool of TDA is the  persistence diagram, and one can use it to compare data sets. One approach to learn on the similarity between two persistence diagrams is to use the Bottleneck and the Wasserstein distances. Another approach is to fit a parametric model for each diagram, and then to compare the model coefficients. We study the behaviour of both distance measures and the RST parametric model. The theoretical behaviour of the distance measures is difficult to be developed, and therefore we study their behaviour numerically. We conclude that the RST model has an advantage over the Bottleneck and the Wasserstein distances in sense that it can give a definite conclusion regarding the similarity between two persistence diagrams. More of that, a great advantage of the RST is its ability to distinguish between two data sets that are geometrically different but topologically are the same, which is impossible to have by the two distance measures.
\end{abstract}


\section{Introduction}
\noindent The term of 'big data' is commonly used for describing a high-dimensional, incomplete and noisy data. That is, it describes a large sample size, or, alternatively, a small sample size but with a large number of measurements (covariates) for each sample unit.
Topological data analysis (TDA) is a powerful tool to handle with big data, where its initial motivation is to study the shape of data. The main tool in TDA is the persistent homology, an adaptation of homology to point cloud data.
Specifically, the persistence diagram is a summarized description of the data shape in terms of topological features such as connected components, holes, voids, etc. (see, for example, \cite{EdelsHarerBook}).
Every point of the persistence diagram is a two-dimensional point, and represents the persistent generator. The first coordinate of such point presents the first filtration level where it appears, the 'birth time', and its second coordinate presents the filtration level where it disappears, the 'death time'.
Hence, given two big data sets, one can use their persistence diagrams to determine if they behave 'the same' or not. Formally, let $F_i$ denote the distribution function of the $i$-th data set, $i=1,2$. Checking if two data sets behave the same is equivalent to testing the null hypothesis $H_0:F_1=F_2$ versus the alternative $H_1: F_1 \ne F_2$. Given the persistence diagram for each data set, the comparison of the two persistence diagrams can be done by using the Bottleneck and the Wasserstein distances between the two persistence diagrams. These distances measure the similarity between the two persistence diagrams (see, for example, \cite{EdelsHarerBook}) by describing the cost of the optimal matching between points of the two diagrams. Note that all the diagonal points are included in the persistence diagrams when computing the optimal matching.
Clearly, as these two distances are smaller and close to zero, the two corresponded data sets are more likely to behave the same, and vice versa. In order to determine which values of these distances are considered small and which are not, one needs to explore the distribution of each distance measure under the hypothesis that the two persistence diagrams are similar. Particularly, this determination is important when the data includes noise, since then it might be that two persistence diagrams are corresponded to the same phenomenon, but due to the noise they are different, which yields positive values of the Bottleneck and Wasserstein distances. The theoretical development of such distribution is difficult, and therefore a simulation study is needed. The simulation study can carried out using a large number of paired persistence diagrams based on samples from the same phenomenon. Alternatively, it can use resampling of data sets from the one original paired data set, or resampling of persistence diagrams from the original one corresponded paired persistence diagram. Other approach for checking similarity between two persistence diagrams is to use statistical inference on persistent homology. This can be nonparametric, that is, without characterize the distribution of topological features, or it can be parametric, that is, based on some parametric model for the points on each persistence diagram. \cite{Robinson} suggested the nonparametric permutation test which is based on a joint loss function and a randomization test.
\cite{Ferri} represented the persistence diagram as the set of complex roots of a polynomial, and then comparison can be performed on the coefficients.
\cite{Fabio} suggested an algebraic representation of persistence diagrams by complex polynomials: far polynomials represent far persistence diagrams, therefore a fast comparison of the coefficient vectors can reduce the size of the database to be classified by the bottleneck distance.
The RST is a parametric model for the points on the persistence diagram suggested by \cite{adleragam2}. Their model involves three parameters that capture the spread of the points on the persistence diagram in terms of nearest neighbors, and another nuisance parameter that involved in the kernel density estimator (KDE) which captures the shape of the persistence diagram. Fitting this model to each of the paired persistence diagram enable to examine the difference between the paired persistence diagrams by examining the difference between the model parameters; If this difference for all the three parameters is simultaneously different from zero, then the two data sets are different, and vice versa.
In this paper we study the distribution of the Bottleneck and Wasserstein distances under some examples, and compare it with the performance of the parametric RST model. The considered examples include two data sets in each one, for which we know ahead they behave the same or not. Section 2 presents the background, describes the two distance measures Bottleneck and Wasserstein, and describes the RST model. Section 3 presents a simulation study comparing the distributions of the Bottleneck and Wasserstein distances with the RST model fitting. Section 4 presents a real data example of weather in Israel, and compares the weather over two different cities in Israel. Section 5 gives a brief summary.

\section{Methods}
\subsection{Setting and Notation}
\noindent Let $\cZ$ be some space, and let $f$ be a smooth real function over $\cZ$. Suppose we observe a sample $\tZ_n=\{Z_1,\dots,Z_n\}$ drawn from a distribution $P$ supported on $\cZ$. Denote by $A_{l}$  the lower-level sets of the form $A_l=\{z\in \cZ : f(z)\leq l\}$. As $l$ varies from $0$ to $\infty$, the topological features of $A_{l}$, such connected components (homology of zero rank, $H_0$), holes (homology of first rank, $H_1$), voids (homology of second rank, $H_2$), etc., change. These features can appear and disappear as $l$ increases. The value of $l$ for which a topological feature appears is called the 'birth time' ($b$), and the value of $l$ for which the topological feature disappeared is called the 'death time' ($d$). Note that $b<d$. By the same way we can look at the upper-level sets $A_u=\{z\in \cZ : f(z)\geq u\}$, and then $d<b$. The collection of the points $(b,d)$ (or $(d,b)$) plotting on two axes is called the persistence diagram. We denote by $N$ the number of points on the persistence diagram that present the same homology.
Usually, the function $f$ is a distance function or a smooth function such as the kernel density estimator (KDE). In this paper we consider the persistence diagram that based on the upper-level sets with $f$ be the KDE. We observe two samples
$\tZ_{n_1}$ and $\tZ_{n_2}$ drawn from the distributions $P_1$ and $P_2$, where $P_1$ may be equal to or differ from $P_2$.
\subsection{Distance Measures}
\noindent For comparing two persistence diagrams, there are two known measures: the bottleneck distance and the Wasserstein distance \cite{EdelsHarerBook}. Let $\tilde{B}_1$ and $\tilde{B}_2$ be two persistence diagrams; the bottleneck distance is defined by
\beqq
W_{\infty}\left(\tilde{B}_{1},\tilde{B}_{2}\right)=\mathop{\inf}\limits_{g:\,\tilde{B}_{1}\to\tilde{B}_{2}}
\mathop{\sup}\limits_{x\in\tilde{B}_{1}}\left\| x-g\left(x\right)\right\|_{\infty},
\eeqq
where the infimum is over all bijections from $\tilde{B}_1$ and $\tilde{B}_2$. That is, the bottleneck distance is the maximum distance between the points of the two persistence diagrams, after minimizing over all possible pairings of the points, including the points on the diagonals.
The $p$-th Wassetrstein distance between $\tilde{B}_1$ and $\tilde{B}_2$ is defined by
\beqq
W_{p}\left(\tilde{B}_{1},\tilde{B}_{2}\right)=\left[\mathop{\inf}\limits_{g:\,\tilde{B}_{1}\to\tilde{B}_{2}}
\mathop{\sup}\limits_{x\in\tilde{B}_{1}}\left\| x-g\left(x\right)\right\|_{\infty}^{p}\right]^{{1\mathord{\left/{\vphantom{1 p}}\right.\kern-\nulldelimiterspace} p}}.
\eeqq
\noindent These two distances are interesting since they are stable \cite{Choen}: a small change of the measured data creates only a small change in the persistence diagram.
As noted in \cite{EdelsHarerBook}, the bottleneck distance is the cruder of the two distances, and the Wasserstein distance is more sensitive to details in the persistence diagram.
For two persistence diagrams that are obtained from similar data set, we expect that these two distances will be close to zero. But, this can depends on the number of points $n$ of the data set, and on the measurement accuracy of the data.
Clearly, if the number of points $n$ is too small, or, if the amount of the noise is too large, it may be no longer true to have a zero distance or a distance that is close to zero. Therefore it is interesting to explore the distribution of these two distance measures as a function of $n$ and the amount of noise in the data. Since it is difficult and may impossible to examine the theoretical distribution of these two measures, we study it via a simulation study.

\subsection{The RST Model}
\noindent The following is a description of the suggested modeling of \cite{adleragam2} to $N$ points with the same rank of homology on the persistence diagram.
Define a new set of $N$ points $\tx_N =\{x_i\}_{i=1}^N$, with $x^{(1)}_i=b_i$ and
 $x^{(2)}_i=d_i-b_i$. That is, $\tx_N$ a set of $N$ points in $\cX=\real\times\real_+$.
The goal is fitting a parametric model for $\tx_N$.
For  $x\in\cX$ and for $k\geq 1$ let
$x^{nn}(k) \in \cX$ be the $k$-th nearest neighbour to $x$, let
\beqq
\mathcal L_{\delta,k}(\tx_N) =  \sum_{x\in\tx_N} \|x-x^{nn}(k) \}.
\eeqq
In addition, based on the sample $\tZ_n$ for a compact subset $\cZ$ of $\mathbb R^D$,
let $\hat f_n$ be the Gaussian kernel density estimator (KDE), given by
\beq
\label{eq:kernel}
\hat f_n(p) \ = \  \frac{1}{n(\sqrt{2\pi}\eta)^D}  \sum_{i=1}^{n} e^{{-\|p-z_i\|^2}/{2\eta^2}},\qquad p\in\mathbb R^D,
\eeq
 where $\eta >0$ is a bandwidth parameter for the Gaussian kernel defining $\hat f_n$.
 \\
Define
\beq
\label{eq:hamiltonian1}
\tilde H_{\Theta}^K(\tx_N)
=\sum_{k=1}^K \theta_k \tilde {\mathcal L}_{k}(\tx_N),
\eeq
where $\Theta=(\theta_1,\dots,\theta_K)$, and $K$ is the cluster size.
\\
The considered likelihood (pseudolikelihood \cite{Besag,chalmond}) is
\beq
\label{eq:pseudo1}
\tilde{L}^ K_{\alpha,\Theta}(\tx_N)  \definedas
\prod _{x\in\tx_N}  f_\Theta \left(x\big|    \cN_{K}(x) \right),
\eeq
 where
$\cN_{K}(x)$ denotes the $K$ nearest neighbours of $x$ in $\tx_N$, and
 \beq
  f_\Theta\left(x\big|    \cN_{K}(x) \right)=\frac{
 (KDE(x))^{\alpha}\times \exp \left(-\tilde{H}^K_{\alpha,\Theta}\left(x\big|   \cN_{K}(x) \right)\right)
  }{
\int _{\real}\int _{\real_+} (KDE(z))^{\alpha} \times \exp \left(-\tilde{H}^K_{\alpha,\Theta}\left(z\big|   \cN_{K}(x) \right)\right)\,dz^{(1)}dz^{(2)},
}
\label{eq:conditionalham1}
\eeq
with
\beqq
\tilde{H}^K_{\alpha,\Theta}\left(x\big|   \cN_{K}(x)\right)=\sum_{k=1}^K \theta_k\tilde{\cL}_{\alpha,k}\left(\cN_{K}(x)\right).
\eeqq
\\
The parameter $\alpha$ is a non-negative nuisance parameter. Note that nearest neighbors captures the closeness relations between the points, and the KDE controlling the shape of the whole points on the persistence diagram.
For considering some values of $K$, the best model can be chosen by the automated statistical procedures such as AIC, BIC, etc.\ (cf.\ \cite{burnham}). The estimation of $\alpha$ is done by the bisection method, where after considerable experimentation, \cite{adleragam2} found that it is enough to take the search (non-negative) range to be $[0,3]$.
\\
Given the value of $\alpha$ that maximizes the log likelihood, one can search for $\Theta$ that maximizes the log likelihood.
Based on the model, we can check if two data sets come from the same distribution by fitting a model to each data set and then comparing the estimates. This can be done by testing the hypothesis $H_{0} :\theta _{j}^{1} =\theta _{j}^{2}$ vs. the alternative hypothesis $H_{1} :\theta _{j}^{1} \ne \theta _{j}^{2}$, for $j=1,2,3$.
 The test statistic is $\hat{\theta} _{j}^{1} -\hat{\theta} _{j}^{2} $, where $\hat{\theta} _{j}$ is the estimated $\theta _{j}$. Based on the asymptotic properties of the pseudo likelihood (\cite{Jensen}, \cite{Billiot}), the $p$-value is obtained by $2\times P\left(Z\ge \left|\frac{\hat{\theta} _{j}^{1} -\hat{\theta} _{j}^{2} }{\sqrt{Var\left(\hat{\theta} _{j}^{1} \right)+Var\left(\hat{\theta} _{j}^{2} \right)} } \right|\right)$ , where $Z$ is the standard normal random variable.

\section{Simulation Study} 
\noindent In this section we compare via examples the behaviour of the Bottleneck and Wasserstein distances, with the performance of the RST model results. Tables A.1-A.4 in the Appendix present the results.
\subsection{The Simulation Design}
\noindent We present six examples, each example contains two samples with size $n$ each one, from two distributions $P_1$ and $P_2$. The distributions $P_1$ and $P_2$ can be divided into two classes: In one class, $P_1$ and $P_2$ are the same, and particularly, have the same topology and geometry. In the second class, $P_1$ and $P_2$ are different, but the difference is geometrically only, whereas topologically the two distributions are the same.
In the first class, we consider three examples of $P_1$ and $P_2$: one circle, two distinct circles, and two concentric circles.
In the second class, we consider four examples of $P_1$ and $P_2$: one circle with two different radii, two different objects of two distinct circles, two different objects of two concentric circles, and two distinct circles vs. of two concentric circles.
Note that all the examples are two-dimensional objects.
The aim is to compare two $H_0$ persistence diagrams at each example, where each persistence diagram is generated by the super-level sets of the fitted kernel density estimator for the data with an arbitrary bandwidth of 0.1. We consider the data to be measured without any noise.
For each example we take $n$ to be 100, 300, 500, 1000, 1500, 2000, 2500, 3000, 20000, 25000, 30000. We generated 1000 data sets from each $P_i$, $i=1,2$, and for each paired persistence diagrams we calculated its Bottleneck and Wasserstein distances, and the fitted RST model for each persistence diagram. By this we got for each example 1000 distances of Bottleneck and Wasserstein, and 1000 differences of the RST model estimates for a given parameter.
We summarized the results of the Bottleneck and Wasserstein distances by calculating their range, interquartile-range (IQR), and the standard deviation (std) of each distance measure over the 1000 distances. This summary is presented in the first two column blocks of Tables A.1-A.4, in a 3-digit accuracy after the decimal point. In addition we present the ratio of the Bottleneck range with the Wasserstein range in the third columns block, and it refers to the results in a 4-digit accuracy after the decimal point. For the RST model, we calculated the $p$-value of each difference for a given parameter, and counted for each pair the number of significant differences from the three differences. Then we calculated the proportion of $k$ significant differences for each parameter over the 1000 pairs, where $k=0,1,2,3$. This proportion is presented in the last column block of Tables A.1-A.4.

\subsection{Examples}
\begin{landscape}
\begin{figure}[h!]
\bc
\includegraphics[width=1.8in, height=1.8in]{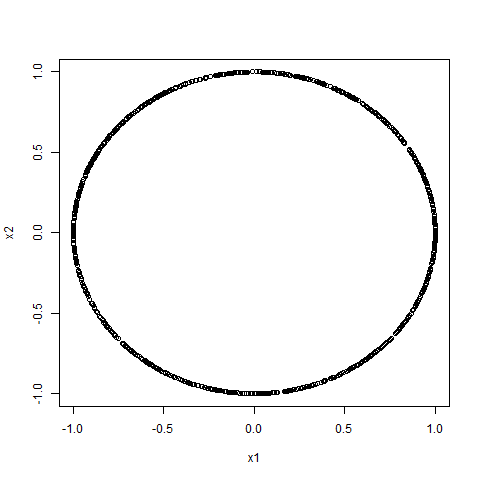}
\includegraphics[width=1.8in, height=1.8in]{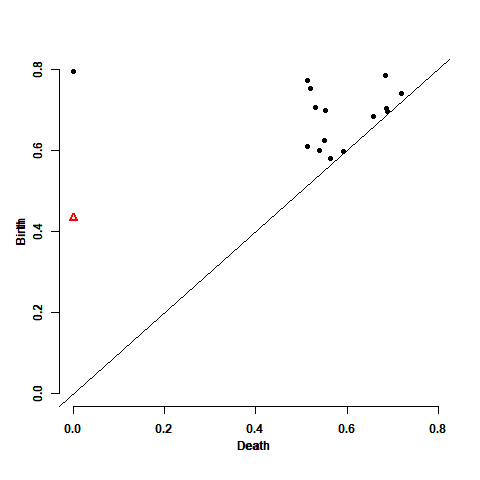}
\includegraphics[width=1.8in, height=1.8in]{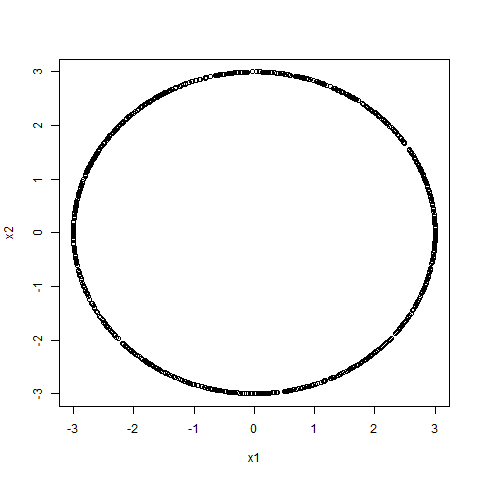}
\includegraphics[width=1.8in, height=1.8in]{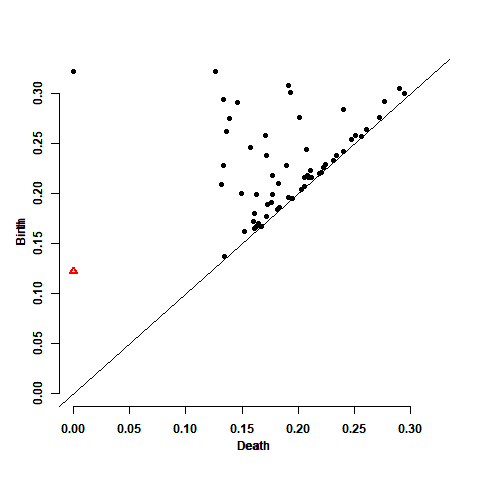}
\includegraphics[width=1.8in, height=1.8in]{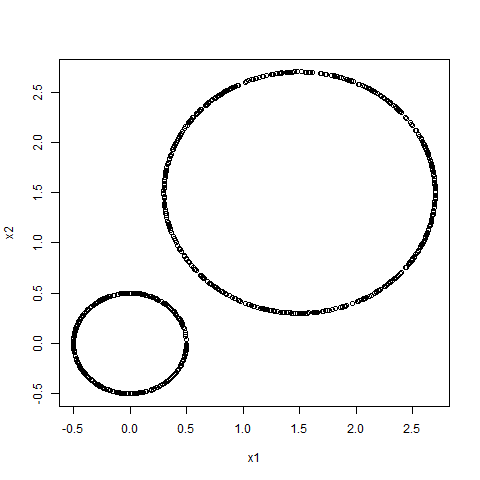}
\includegraphics[width=1.8in, height=1.8in]{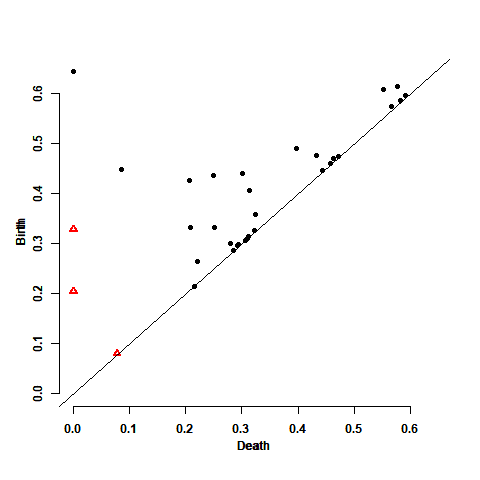}
\includegraphics[width=1.8in, height=1.8in]{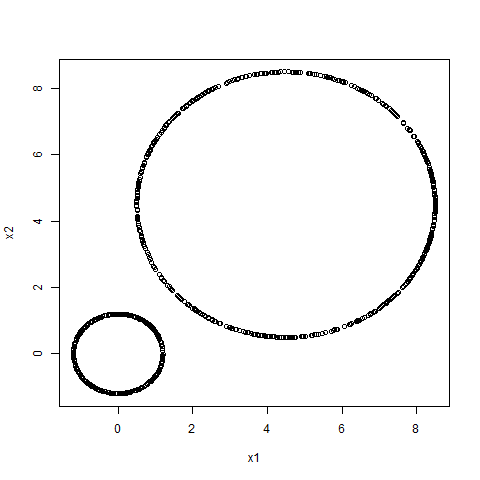}
\includegraphics[width=1.8in, height=1.8in]{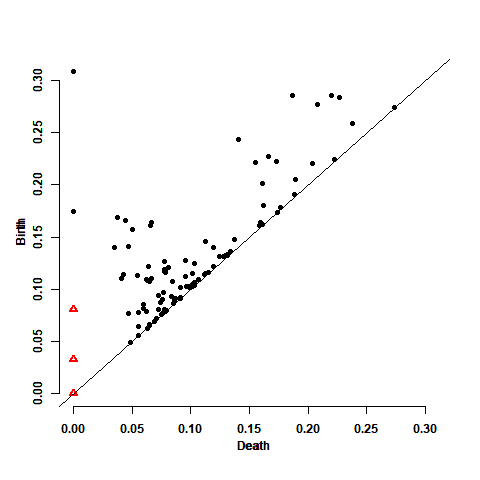}
\includegraphics[width=1.8in, height=1.8in]{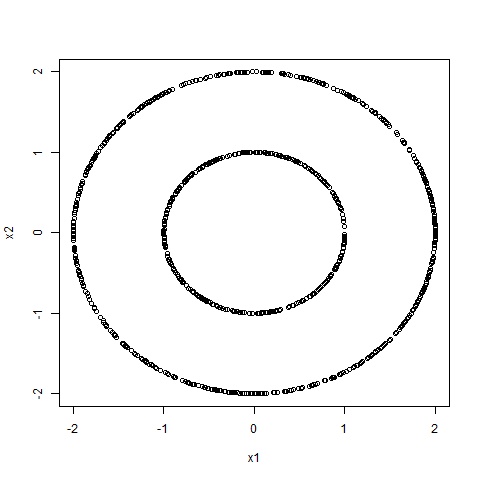}
\includegraphics[width=1.8in, height=1.8in]{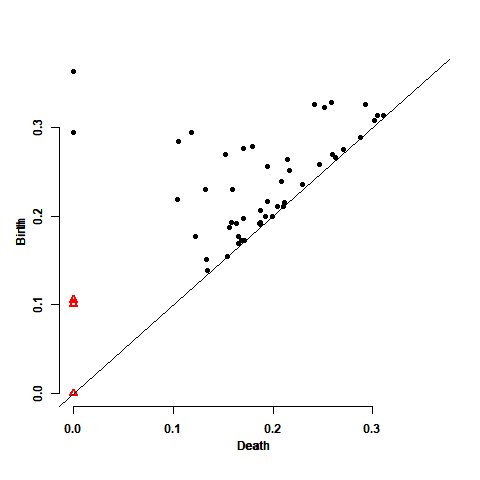}
\includegraphics[width=1.8in, height=1.8in]{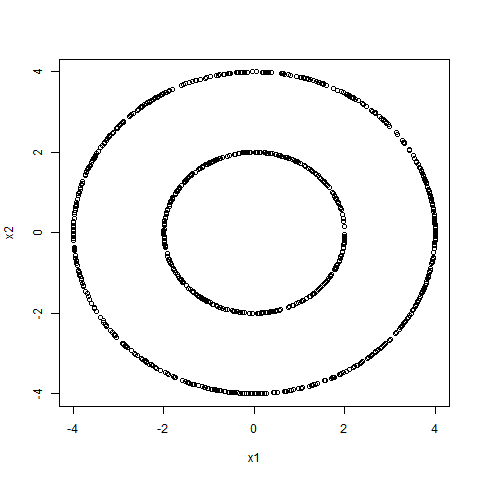}
\includegraphics[width=1.8in, height=1.8in]{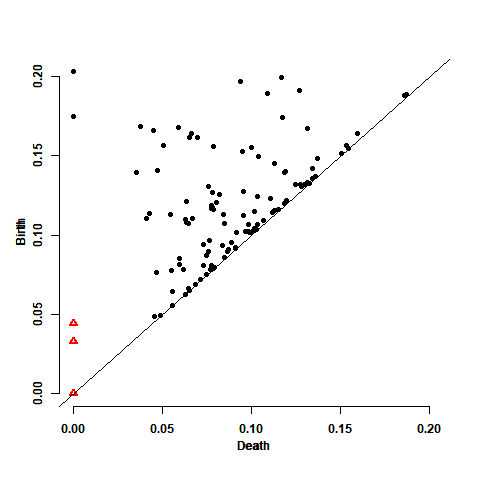}
\ec
\caption{\footnotesize
 The examples described here, from left to right, are: a unit circle, a circle with radius $r=3$, two distinct circles with a distance of 1.5 for each point, two distinct circles with a distance of 4.5 for each point, two concentric circles with $r_1=1, r_2=2$, and two concentric circles with $r_1=2, r_2=4$. For each example, the first plot describes the data, and the second plot describes the corresponded persistence diagram for its upper level sets. Black circles are connected components ($H_0$ persistence points), red triangles are holes ($H_1$ points), blue diamonds are voids ($H_2$ points). Birth times are on the vertical axis. See the text for more details.}
\label{fig:circle}
\end{figure}
\end{landscape}

\noindent The examples we consider are:
\\
1. \emph{One Circle} - A comparison of two samples drawn from a circle with radius $r=1$ (the unit circle).
 \\
2. \emph{Two Circles with Different Radii} -A comparison of a sample from a unit circle with a sample from a circle with radius $r=3$.
\\
3. \emph{Two Distinct Circles}- This object contains one circle with radius $r_1=0.5$, and a second circle with $r_2=1.2$. The distance between these two circles is 1.5 for each point. The number of points of the smaller circle and the larger circle is $0.4n$ and $0.6n$, respectively.
\\
4. \emph{Different Two Distinct Circles} - A comparison of two samples where each one is taken from a different object of two distinct circles. The first object is as in the previous example. The second object includes one circle with radius $r_1=1.2$, and the second circle with radius $r_2=4$. The distance between these two circles is 4.5 for each point. The number of points of the smaller circle and the larger circle is $0.4n$ and $0.6n$, respectively.
\\
5. \emph{Two Concentric Circles} - The object of concentric circles contains two circles: One circle has radius $r_1=1$, and the second circle has radius $r_2=2$. The number of points of the smaller circle and the larger circle is $0.4n$ and $0.6n$, respectively. The both circles together obtain a smaller circle inside a larger one.
\\
6. \emph{Different Two Concentric Circles} - A comparison of two samples where each one is taken from a different object of two concentric circles. The first object is as in the previous example. The second object includes one circle with radius $r_1=2$, and the second circle with radius $r_2=4$. The number of points of the smaller circle and the larger circle is  $0.4n$ and $0.6n$, respectively.
\\
7. \emph{Two Distinct Circles Vs. Two Concentric Circles} - The object of distinct circles is  the same as in Example 3. The object of concentric circles is the same as in Example 5.
\\
Each example is described in Figure\ \ref{fig:circle} based on a sample of $n=1000$. The first plot of each example describes the object sample, and to its right we present the corresponding persistence diagram. The black circles indicating connected components ($H_0$ persistence), the red triangles corresponding to holes ($H_1$), and the blue diamonds corresponding to voids ($H_2$).

\subsection{The Calculation of the $P$-Value}
\noindent As was mentioned above, the asymptotic distribution of the difference $\hat{\theta} _{j}^{1} -\hat{\theta} _{j}^{2} $ is normal, and then one can use the regular $p$-value.
Let's denote by $\Delta_{j}$ the difference $\hat{\theta} _{j}^{1} -\hat{\theta} _{j}^{2}$. In the examples we considered, we checked if the fitted distribution of $\Delta_{j}$ is normal or not, using the Kolmogorv-Smirnov test: If the distribution is indeed normal, then we used the $p$-value based on the asymptotic properties mentioned above. Otherwise, we used the empirical two-sides $p$-value described in \cite{GEORGE}, as follows: Let's denote by $\delta$ the observed value of $\Delta$, and let $\delta^{*}$ be such that $f_0(\delta^{*})=f_0(\delta)$, where $f_0$ denotes the empirical distribution of the estimated $\Delta$. Then, the observed $p$-value is given by $p(t)=min[1-F(t)+F(t^{*}),1-F_0(t^{*})+F_0(t)]$, where $F(t)$ is the empirical cumulative distribution of $\Delta$ at point $t$. For both situations, we used finally the Bonfferoni correction for multiple comparisons, for significance level of $\alpha =0.05$ .

\subsection{Results}
\noindent For the first class of examples which are topologically and geometrically the same, the Bottleneck and Wasserstein distances become smaller toward zero as $n$ increases, as expected. The rate of the tendency toward zero is faster when the data complexity is smaller, and vice versa. Interesting, when the data complexity is smaller (for example, in one circle, or two distinct circles), the Bottleneck distance goes slower to zero as $n$ increases relative to the Wasserstein distance. That is, the ratio between the Bottleneck and the Wasserstein distances increases as $n$ increases. Particularly, in the considered examples, this ratio can ranged from 11 to 22 in large $n$. But, when the data complexity is higher (for example in the two concentric circles), the Bottleneck distance can go faster to zero as $n$ increases relative to the Wasserstein distance, and the ratio between them can ranged from 0.36 to 1. The specific values of the Bottleneck distance, based on the considered examples and for large $n$, ranged from 0.005 to 0.048, and of the Wasserstein distance, from 0.001 to 0.023. Particularly, this means that the two distance measures are not necessarily negligible for large $n$.
From the other hand, the RST model fitting for each of the two persistence diagrams is resulted in a large probability (between 0.9 to 0.95) of zero difference over the three model parameters. In addition, as can be seen in Tables A.1-A.4, the RST model has an advantage over the two distance measures especially in the smaller sample sizes, where the distance measures are relative large although the two samples behave the same (topologically and geometrically). That is, when two data sets are topologically and geometrically the same, the RST gives a more definite conclusion regarding the similarity between their corresponded persistence diagrams relative to the conclusion obtained by the distance measures.
For the second class of examples for which are topologically the same but geometrically different, the distance measures are relative large and far from zero (although become smaller as $n$ increases), which implies that the two persistence diagrams, and their corresponded data sets as well, are different. In this case, the Bottleneck distance is larger than the Wasserstein distance only for $n\ge 1000$, and the ratio between these two measures is around 1-5 (higher for large $n$). The specific values of the Bottleneck distance, based on the considered examples and for large $n$, ranged from 0.068 to 0.354, and of the Wasserstein distance, from 0.014 to 0.149.
Regarding the RST model fitting, we can see an interesting result: the main probability mass is on one or two different parameters. This means that there is some difference between the two fitted models, but this difference is not full (a full difference is expressed in a difference of all the three model parameters).
That is, the RST model captures the geometric difference, but also the equal topological properties, whereas the distance measures capture the geometric difference only.

\section{Real Example}
\noindent The following example contains a real data of Israel weather in the period $1/6/2018-1/6/2019$. The measurements are in time-resolution of 10 minutes. The considered variables are: (i) Temperature (Temp), measured in degrees Celsius and presents the average temperature in the last 10 minutes (ii) Humidity, measured in percent (iii) Rain, measured in mm and presented the cumulative rain in 10 minutes. Two cities are compared: Jerusalem and Eilat. The number of measurements in Jerusalem were 52669, and 51656 in Eilat. These two cities have a different weather over the year: Eilat has a hot desert climate with hot, dry summers and warm and almost rainless winters; Jerusalem is characterized by a hot-summer Mediterranean climate, with hot, dry summers, and mild, wet winters. Classifying the data into two series of cold and hot days (low and high temperature), with the corresponded months of November to March, and April to October, respectively, yields the summary in Table 1. This summary includes the range of each variable and its median over the whole data. We can see that Jerusalem is characterized by lower temperature and higher humidity relative to Eilat in both cold and hot days.

\begin{center}
\fontsize{7.4}{1.16}\selectfont
\begin{tabular}{l|l|cc|cc}

\mc{6}{c}{\bf{Table 1.} Weather in Israel}\\
\\
\\
\\
\\
\\
\\
\\
\\
&&\textbf{Cold} && \textbf{Hot}\\
\\
\\
\\
\\
\\

&&Jerusalem & Eilat& Jerusalem & Eilat\\
\\
\\
\\
\\
\\
\\
\textbf{Temp}& range& 0.30-16.80 &7.40-25.30 &16.90-37.60 &25.40-44.50 \\
\\
\\
\\
\\
\\
& median&10.70&19.30&22.20&31.10\\
\\
\\
\\
\\
\\
\\
\\
\\
\\
\\
\textbf{Humidity}& range&14-100 &10-96&6-100&6-79\\
\\
\\
\\
\\
\\
    & median&78.00&43.00&55.00&29.00\\
\\
\\
\\
\\
\\
\\
\\
\\
\\
\\
\textbf{Rain}& range&0-3.4  &0-2.7 &0-0.7 &0-0.3\\
\\
\\
\\
\\
\\
    & median&0.00&0.00&0.00&0.00\\
\\
\\
\\
\\
\\
\end{tabular}
\end{center}
\footnotesize{Weather in Israel in $1/6/2018-1/6/2019$, time-resolution of 10 minutes.}
\normalsize

\begin{figure}[h!]
\bc
\includegraphics[width=2.0in, height=2.0in]{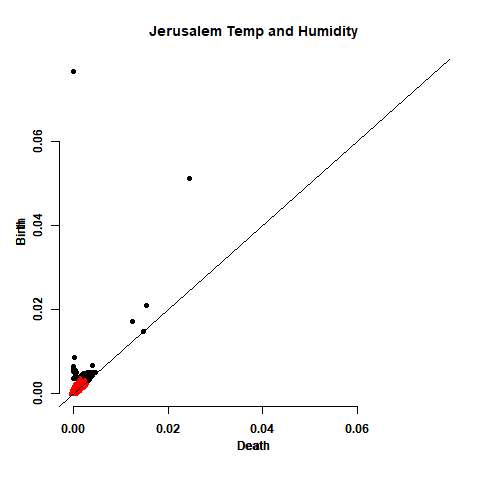} 
\includegraphics[width=2.0in, height=2.0in]{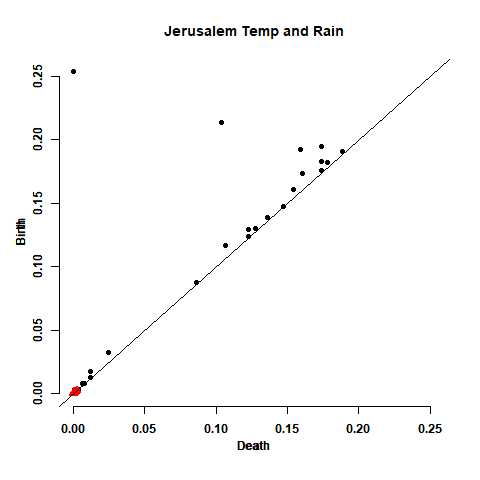} 
\includegraphics[width=2.0in, height=2.0in]{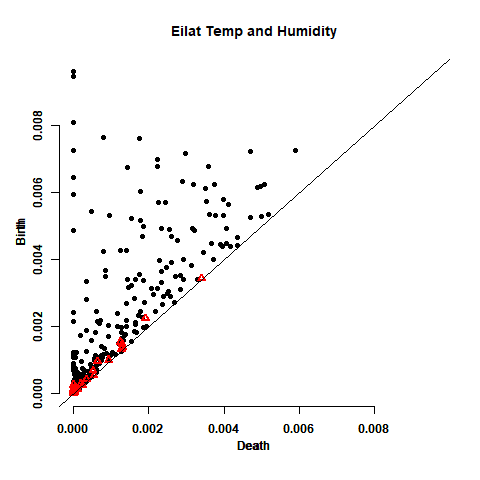}
\includegraphics[width=2.0in, height=2.0in]{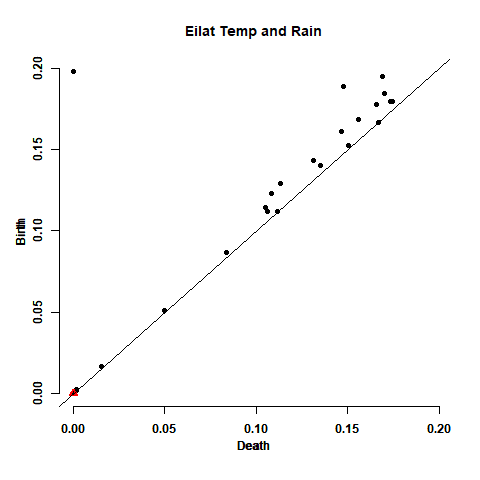}
\ec
\caption{\footnotesize
 Persistence diagrams for weather data based on KDE and the upper level sets. Black circles are $H_0$ persistence points,  red triangles are  $H_1$ points. Birth times are on the vertical axis.}
\label{fig:weather}
\end{figure}

\noindent We want to compare the weather of the two cities by means of TDA and persistence diagram. First we compared the two cities based on the whole data, with the combinations of Temp-Humidity and Temp-Rain. For each combination we calculated the persistence diagram based on the KDE (with a bandwidth of 0.1). The obtained diagrams are presented in
Fig.\ \ref{fig:weather}.
\newpage

\noindent Then we fitted the RST model for each persistence diagram for the $H_0$ points, and compared the coefficients of the model over the two cities. For the $H_1$ points, it was impossible to fit the model because many points were with negligible values. The results are presented in Table 2.

\begin{center}
\fontsize{7.4}{1.16}\selectfont
\begin{tabular}{l|lccc|ccc}

\mc{8}{c}{\bf{Table 2.} Weather in Israel, RST and Distance Measures}\\
\\
\\
\\
\\
\\
\\
\\
\\
&&&\textbf{RST} &&& \textbf{Distance Measures}\\
\\
\\
\\
\\
\\
&\boldmath{$\theta_1$} &\boldmath{$\theta_2$} & \boldmath{$\theta_3$} & \textbf{Difference} & \textbf{Bottleneck} & \textbf{Wasserstein}&\textbf{Difference}\\
\\
\\
\\
\\
\\
\textbf{Temp-Humidity}\\
\\
\\
\\
\\
\\
\textbf{$H_0$}& 0.759&0.944&0.119& No&0.038&0.002&No \\

\\
\\
\\
\\
\\
\\
\\
\\
\\
\\
\textbf{$H_1$}& &&&&0.001&3.33e-06&No\\
\\
\\
\\
\\
\\
\\
\\
\\
\\
\\
\textbf{Temp-Rain}\\
\\
\\
\\
\\
\\
\textbf{$H_0$}& 0.275&0.263&0.207& No&0.056&0.006&No \\

\\
\\
\\
\\
\\
\\
\\
\\
\\
\\
\textbf{$H_1$}& &&&&3.77e-04&2.09e-07&No\\

\end{tabular}
\end{center}
\footnotesize{Weather in Israel in $1/6/2018-1/6/2019$, time-resolution of 10 minutes.}
\normalsize

\noindent Bases on the RST results, there is no difference between the two cities in the level of $H_0$. The results of the Wasserstein distance in the levels of $H_0$ and $H_1$ points were also terminated with no difference between the two cities. Regarding the Bottleneck distance, it is pretty clear that Bottleneck result agrees with the Wasserstain result in the level of the $H_1$ points. For the $H_0$ points, the Bottleneck distance does not close to zero, but note that its value is 19 and 9 times the Wasserstein distance for Temp-Humidity and Temp-Rain, receptively. That is, according to the simulation study, this together with negligible value of the Wasserstein distance, indicate on no difference between the two cities. The conclusion is then that there is no difference in weather of Jerusalem and Eilat once we compare the whole data set. As a second step, we compared the above combinations under low and high temperatures separately. That is, a comparison between cold and hot days. The number of the non-missing measurements in the cold days for Jerusalem and Eilat was 24590 and 23426, respectively. The number of the non-missing measurements in the hot days for Jerusalem and Eilat was 28060 and 21374, respectively. The results are presented in Table 3. For the $H_0$ points, the RST results show a difference between the two cities in their temperature, humidity and rain for the cold days, and a difference in temperature and humidity only but not in the rain for the hot days, as we expect to get. For the cold days, it is unclear how to interpret the results of the Bottleneck and Wasserstein distances; the result of the Wasserstien is not negligible, and, in addition, the Bottleneck result is about 10 and 7 the Wasserstien result for Temp-Humidity and Temp-Rain, respectively. But for the $H_1$ points, it is clear that both distances terminated with no difference for both combinations. For the hot days, the value of Wasserstein is negligible and the Bottleneck is 6 times the Wasserstein, therefore there is no difference between the two cities for Temp-Humidity, For Temp-Rain it is again unclear if there is a difference or not, since the Wasserstien is not negligible and the Bottleneck is about 9 times the Wasserstein value.

\begin{center}
\fontsize{7.4}{1.16}\selectfont
\begin{tabular}{l|lccc|ccc}

\mc{8}{c}{\bf{Table 3.} Cold and Hot Weather in Israel, RST and Distance Measures}\\
\\
\\
\\
\\
\\
\\
\\
\\
\rowcolor{lightgray} \mc{8}{c} {\bf \emph{Cold Weather in Israel}} \\
\\
\\
\\
\\
\\
\\
\\
\\
&&\textbf{RST} &&&& \textbf{Distance Measures}\\
\\
\\
\\
\\
\\
&\boldmath{$\theta_1$} &\boldmath{$\theta_2$} & \boldmath{$\theta_3$} & \textbf{Difference} & \textbf{Bottleneck} & \textbf{Wasserstein}&\textbf{Difference}\\
\\
\\
\\
\\
\\
\textbf{Temp-Humidity}\\
\\
\\
\\
\\
\\
\textbf{$H_0$}& 4.51e-04&0.010&0.006& Yes&0.081&0.008&No \\

\\
\\
\\
\\
\\
\\
\\
\\
\\
\\
\textbf{$H_1$}& &&&&3.05e-04&1.97e-07&No\\
\\
\\
\\
\\
\\
\\
\\
\\
\\
\\
\textbf{Temp-Rain}\\
\\
\\
\\
\\
\\
\textbf{$H_0$}& 0.006&0.011&0.008& Yes&0.077&0.011&Yes/No \\

\\
\\
\\
\\
\\
\\
\\
\\
\\
\\
\textbf{$H_1$}& &&&&0.001&9.05e-07&No\\
\\
\\
\\
\\
\\
\\
\\
\\
\\
\\
\rowcolor{lightgray}\mc{8}{c} {\bf \emph{Hot Weather in Israel}} \\
\\
\\
\\
\\
\\
\\
\\
\\
&&\textbf{RST} &&&& \textbf{Distance Measures}\\
\\
\\
\\
\\
\\
&\boldmath{$\theta_1$} &\boldmath{$\theta_2$} & \boldmath{$\theta_3$} & \textbf{Difference} & \textbf{Bottleneck} & \textbf{Wasserstein}&\textbf{Difference}\\
\\
\\
\\
\\
\\
%
%
%
\textbf{Temp-Humidity}\\
\\
\\
\\
\\
\\
\textbf{$H_0$}& 2.98E-12&3.82e-11&8.57E-08& Yes&0.006&0.001&No \\

\\
\\
\\
\\
\\
\\
\\
\\
\\
\\
\textbf{$H_1$}&5.37E-10 &1.86e-07&2.89E-07&Yes&0.001&1.26e-05&No\\
\\
\\
\\
\\
\\
\\
\\
\\
\\
\\
\textbf{Temp-Rain}\\
\\
\\
\\
\\
\\
\textbf{$H_0$}& 0.714&0.839&0.993& No&0.094&0.011&Yes/No \\

\\
\\
\\
\\
\\
\\
\\
\\
\\
\\

\end{tabular}
\end{center}
\footnotesize{Each value in the RST columns block is the $p$-value of the difference between Jerusalem and Eilat for a given RST coefficient. Each value in the Distance Measures columns block is the value of the distance between the persistence diagrams of Jerusalem and Eilat.}
\normalsize
\section{Summary}
\noindent In this paper we studied the performance of the two known distance measures, the Bottleneck and the Wasserstein distances, for addressing the similarity of two persistence diagrams. We compared it with the results of the RST parametric model. We have found that for two data sets that are topologically and geometrically the same, the value of the distance measures is not always negligible for large data size, and therefore it is not always clear which conclusion to make. But the RST can give a more definite conclusion. In addition, the RST has a great advantage on the two distance measures when the corresponded two data sets are topologically the same but geometrically different; the RST model can capture the both properties of topology and geometry, whereas the distance measures refer to the geometrically difference only. Therefore, for comparing two persistence diagrams, it is recommended to use the RST model in addition to the above standard distance measures.


\newpage
\section*{Appendix}
%

\begin{center}
\fontsize{6.4}{1.16}\selectfont
\begin{tabular}{l|lcc|ccc|cc|cccc}

\mc{11}{c}{\bf{Table A.1.} Sampling From a Circle} \\
\\
\\
\\
\\
\\
\\
\\
\\
\\
\\
\\
\\
\\
\\
\boldmath{$n$$^a$} && \textbf{Bottleneck$^b$}&&& \textbf{ Wasserstein$^c$} &&\textbf{ Ratio$^d$}&&&\textbf{ RST $^e$}\\
\\
\\
\\
\\
\\
\\
\\
\\
& \textbf{Range $^f$}& \textbf{IQR $^g$} & \textbf{Std $^h$}& \textbf{Range}& \textbf{IQR} & \textbf{Std}& \textbf{min}&\textbf{max}& \textbf{0}&\textbf{1}&\textbf{2}&\textbf{3}\\
\\
\\
\\
\\
\\
\\
\\
\\
\\
\\
\\
\\
\\
\\
\\
\\
\mc{10}{c} {\bf Two Samples of a Unit Circle $^i$} \\
\\
\\
\\
\\
\\
\\
\\
\\
\\
\\
\textbf{100} &[0.094,0.964]	&[0.188,0.293]&	0.103&	[0.045,1.041]&[0.136,0.262]&	0.115&2.11&0.93&	0.963&	0.023&	0.013&	0.001\\
\\
\\
\\

\textbf{300}&[0.044,0.461]&	[0.111,0.166]&	0.053&[0.009,0.296]&[0.046,0.085]&	0.035&4.94&1.56&0.940&	0.055&	0.004&	0.001\\
\\
\\
\\
\\

\textbf{500}& [0.037,0.333]&[0.085,0.126]&	0.038&	[0.008,0.167]&	[0.028,0.050]&	0.020&4.53&1.99&0.953&	0.041&	0.006&0\\
\\
\\
\\

\textbf{1000}& [0.029,0.208]&[0.0610.091]&	0.026&[0.005,0.070]&[0.015,0.026]&	0.009&	5.65&2.95&0.912&0.085&0.003&	0\\
\\
\\
\\
\textbf{1500}& [0.028,0.185]&[0.050,0.073]&	0.020&[0.004,0.057]&[0.010,0.017]&	0.006&8.09&3.24&0.908&	0.087&	0.005&	0\\
\\
\\
\\
\textbf{2000}& [0.020,0.214]&[0.044,0.064]&0.018&[0.003,0.058]&[0.008,0.014]&0.005&6.39&3.72&0.947&	0.051&	0.002&	0\\
\\
\\
\\
\textbf{2500}& [0.023,0.131]&[0.039,0.056]&0.016&[0.003,0.026]&[0.007,0.011]&0.004&7.57&5.00&0.921&	0.073&	0.006&	0\\
\\
\\
\\
\textbf{3000}& [0.020,0.147]&[0.036,0.052]&0.015&[0.003,0.032]&[0.006,0.010]&0.003&7.80&4.56&0.941&	0.056&	0.003&	0\\
\\
\\
\\
\textbf{20000}& [0.010,0.048]&[0.016,0.023]&0.006&[0.001,0.005]&[0.002,0.002]&7e-04&16.17&10.41&0.934&0.058&	0.007&0.001\\
\\
\\
\\
\textbf{25000}& [0.009,0.045]&[0.015,0.020]&0.005&[0.001,0.005]&[0.001,0.002]&5e-04&14.50&10.00&0.946&0.046&0.008&0\\
\\
\\
\\
\textbf{30000}& [0.009,0.048]&[0.014,0.019]&0.004&[4e-04,4e-03]&[0.001,0.002]&5e-04&21.75&11.00&0.909&0.083&0.008&0\\
\\
\\
\\
\\
\\
\\
\\
\\
\\
\\
\\
\\
\\
\\
\\
\\
\\

\mc{10}{c} {\bf Two Samples of Circles with Different Radii $^j$} \\
\\
\\
\\
\\
\\
\\
\\
\\
\\
\\
\\
\\
\\
\\

\textbf{100} &[0.257,1.012]	&[0.441,0.654]&	0.140&[0.612,2.080]&[1.036,1.398]&0.252& 0.42&0.49&0.030&	0.694&0.272&0.004\\
\\
\\
\\

\textbf{300}&[0.213,0.712]&	[0.434,0.512]&0.068&[0.327,0.948]&[0.544,0.685]&0.102&0.65&0.75&0.069&0.646&0.283&0.002\\
\\
\\
\\
\\

\textbf{500}& [0.232,0.581]&[0.425,0.471]&0.041&[0.253,0.684]&[0.409,0.497]&0.063&0.92&0.85&0.049&0.603&0.342&0.006\\
\\
\\
\\

\textbf{1000}& [0.338,0.492]&[0.394,0.422]&	0.022&[0.194,0.454]&[0.291,0.331]&	0.031&1.74&1.08&0.053&	0.592&0.351&0.004\\
\\
\\
\\
\textbf{1500}& [0.337,0.473]&[0.379,0.402]&	0.018&[0.177,0.356]&[0.245,0.272]&0.022&1.90&1.33&0.039&0.513&0.442&	0.006\\
\\
\\
\\
\textbf{2000}& [0.326,0.482]&[0.371,0.391]&0.016&[0.159,0.319]&[0.219,0.241]&0.017&	2.05&1.51&0.022&0.527&0.440&0.011\\
\\
\\
\\
\textbf{2500}& [0.345,0.434]&[0.365,0.382]&0.014&[0.174,0.271]&[0.202,0.221]&0.015&1.98&1.60&0.016&0.499&0.479&0.006\\
\\
\\
\\
\textbf{3000}& [0.340,0.431]&[0.361,0.377]&0.013&[0.169,0.246]&[0.192,0.208]&0.012&2.01&1.75&0.014&0.451&0.528&0.007\\
\\
\\
\\
\textbf{20000}& [0.328,0.357]&[0.334,0.340]&0.005&[0.130,0.154]&[0.137,0.142]&0.004&2.52&2.32&0.165&0.809&0.024&0.002\\
\\
\\
\\
\textbf{25000}& [0.326,0.351]&[0.332,0.338]&0.004&[0.128,0.148]&[0.134,0.139]&0.003&2.54&2.37&0.132&0.840&0.027&0.001\\
\\
\\
\\
\textbf{30000}& [0.325,0.354]&[0.331,0.336]&0.004&[0.126,0.149]&[0.132,0.136]&0.003&2.59&2.37&0.002&0.967&0.029&0.002\\

\end{tabular}
\end{center}
\footnotesize{$^a$ Sample size. $^b$ Bottleneck distance between 1000 pairs of $H_0$ persistence diagrams corresponded to 1000 paired samples. $^c$ Wasserstein distance between 1000 pairs of $H_0$ persistence diagrams corresponded to 1000 paired samples. $^d$ Ratio of the Bottleneck range to the Wasserstein range; ratio of the lowest edge ("min"), and ratio of the highest edge("max"). $^e$ Proportion of 0,1,2,3 significant differences of the RST parameters, over 1000 fitted pairs models for 1000 pairs of $H_0$ persistence diagrams. $^f$ Range of 1000 Bottleneck distances. $^g$ Interquartile range of 1000 Bottleneck distances. $^h$ Standard deviation of 1000 Bottleneck distances. $^i$ Two samples from a circle with radius $r=1$, each with $n$ points. $^j$ Two samples from a circle with radius $r=1$ and $r=3$, respectively, each with $n$ points.}
\normalsize

\normalsize
\newpage
\begin{center}
\fontsize{6.4}{1.16}\selectfont
\begin{tabular}{l|lcc|ccc|cc|cccc}
\mc{10}{c}{\bf{Table A.2.} Sampling From Two Distinct Circles} \\
\\
\\
\\
\\
\\
\\
\\
\\
\\
\\
\\
\\
\\
\\
\boldmath{$n$$^a$} && \textbf{Bottleneck$^b$}&&& \textbf{ Wasserstein$^c$} &&\textbf{ Ratio$^d$}&&&\textbf{ RST $^e$}\\
\\
\\
\\
\\
\\
\\
\\
\\
& \textbf{Range $^f$}& \textbf{IQR $^g$} & \textbf{Std $^h$}& \textbf{Range}& \textbf{IQR} & \textbf{Std}&  \textbf{min}&\textbf{max}& \textbf{0}&\textbf{1}&\textbf{2}&\textbf{3}\\
\\
\\
\\
\\
\\
\\
\\
\\
\\
\\
\\
\\
\\
\\
\\
\\
\mc{10}{c} {\bf Two Samples of Two Distinct Circles $^i$} \\
\\
\\
\\
\\
\\
\\
\\
\\
\\
\\
\\
\\
\\
\\
\\
\\
\textbf{100} &[0.095,0.718]	&[0.159,0.251]&0.087&[0.053,0.626]&[0.127,0.212]&0.080&	1.80&1.15&0.949&	0.051&	0&	0\\
\\
\\
\\

\textbf{300}&[0.051,0.352]&	[0.103,0.161]&0.049&[0.020,0.232]&[0.052,0.084]&0.028&2.55&1.52&0.924&0.075&0.001&0\\
\\
\\
\\
\\

\textbf{500}& [0.046,0.368]&[0.081,0.124]&	0.038&[0.012,0.182]&[0.032,0.054]&0.018&3.88&2.02&0.938&0.056&0.006&0\\
\\
\\
\\

\textbf{1000}& [0.034,0.269]&[0.058,0.088]&0.026&[0.007,0.091]&[0.016,0.026]&	0.009&5.04&2.97&0.937&0.056&0.007&	0\\
\\
\\
\\
\textbf{1500}& [0.028,0.163]&[0.045,0.071]&	0.020&[0.005,0.045]&[0.011,0.018]&	0.006&5.79&3.60&0.913&0.071&0.015&	0.001\\
\\
\\
\\
\textbf{2000}& [0.023,0.142]&[0.042,0.060]&0.016&[0.004,0.032]&[0.009,0.013]&0.004&5.63&4.51&0.912&0.076&	0.012&	0\\
\\
\\
\\
\textbf{2500}& [0.020,0.122]&[0.037,0.054]&0.014&[0.003,0.023]&[0.007,0.011]&0.003&6.00&5.29&0.912&0.073&0.014&0.001\\
\\
\\
\\
\textbf{3000}& [0.017,0.123]&[0.034,0.048]&0.013&[0.002,0.025]&[0.006,0.009]&0.003&9.16&5.02&0.918&	0.068&0.014&0\\
\\
\\
\\
\textbf{20000}& [0.008,0.048]&[0.014,0.019]&0.005&[0.001,0.004]&[0.001,0.002]&4e-04&13.67&13.03&0.899&0.084&0.017&0\\
\\
\\
\\
\textbf{25000}& [0.008,0.037]&[0.012,0.017]&0.004&[0.001,0.003]&[0.001,0.002]&4e-04&13.00&12.83&0.914&0.060&0.026&0\\
\\
\\
\\
\textbf{30000}& [0.007,0.033]&[0.011,0.016]&0.004&[0.001,0.003]&[0.001,0.001]&3e-04&13.60&12.15&0.887&0.082&0.031&0\\
\\
\\
\\
\\
\\
\\
\\
\\
\\
\\
\\
\\
\\
\\
\\
\\
\\

\mc{10}{c} {\bf Two Samples of Different Two Distinct Circles $^j$} \\
\\
\\
\\
\\
\\
\\
\\
\\
\\
\\
\\
\\
\\
\\

\textbf{100} &[0.168,0.841]	&[0.302,0.474]&0.123&[0.390,1.573]&[0.640,0.878]&0.177&	0.43&0.53&0.562&	0.425
&0.013&0\\
\\
\\
\\

\textbf{300}&[0.140,0.553]&	[0.278,0.394]&0.077&[0.231,0.714]&[0.385,0.482]&0.073&0.60&0.77&0.001&0.365&0.623&0.011\\
\\
\\
\\
\\

\textbf{500}& [0.139,0.459]&[0.289,0.366]&0.056&[0.192,0.490]&[0.296,0.356]&0.046&0.73&0.94&0.001&0.277&0.713&0.009\\
\\
\\
\\

\textbf{1000}& [0.203,0.431]&[0.292,0.335]&0.034&[0.140,0.326]&[0.210,0.248]&0.028&1.45&1.32&0&	0.292&0.698&0.010\\
\\
\\
\\
\textbf{1500}& [0.201,0.387]&[0.292,0.322]&	0.025&[0.115,0.257]&[0.178,0.203]&	0.019&1.76&1.51&0&0.300&0.690&0.010\\
\\
\\
\\
\textbf{2000}& [0.228,0.363]&[0.291,0.312]&0.019&[0.114,0.214]&[0.160,0.178]&0.015&2.00&1.70&0.001&0.294&0.693&0.012\\
\\
\\
\\
\textbf{2500}& [0.236,0.358]&[0.289,0.308]&0.015&[0.116,0.200]&[0.148,0.164]&0.013&2.03&1.78&0&	0.274&0.714&0.012\\
\\
\\
\\
\textbf{3000}& [0.239,0.345]&[0.287,0.302]&0.013&[0.109,0.205]&[0.138,0.152]&0.011&2.19&1.68&0&	0.269&0.721&0.010\\
\\
\\
\\
\textbf{20000}& [0.261,0.289]&[0.268,0.274]&0.004&[0.088,0.108]&[0.095,0.099]&0.003&2.98&2.67&0.001&0.305&0.685&0.009\\
\\
\\
\\
\textbf{25000}& [0.261,0.284]&[0.267,0.271]&0.004&[0.087,0.104]&[0.093,0.096]&0.003&3.02&2.72&0.001&0.323&0.667&0.009\\
\\
\\
\\
\textbf{30000}& [0.261,0.281]&[0.266,0.271]&0.003&[0.086,0.103]&[0.092,0.095]&0.003&3.03&2.74&0.302&0.687&0.011&0\\
\\
\\
\\
\end{tabular}
\end{center}
\footnotesize{$^a$ Sample size. $^b$ Bottleneck distance between 1000 pairs of $H_0$ persistence diagrams corresponded to 1000 paired samples. $^c$ Wasserstein distance between 1000 pairs of $H_0$ persistence diagrams corresponded to 1000 paired samples. $^d$ Ratio of the Bottleneck range to the Wasserstein range; ratio of the lowest edge ("min"), and ratio of the highest edge("max"). $^e$ Proportion of 0,1,2,3 significant differences of the RST parameters, over 1000 pairs models fitted to 1000 pairs of $H_0$ persistence diagrams. $^f$ Range of 1000 Bottleneck distances. $^g$ Interquartile range of 1000 Bottleneck distances. $^h$ Standard deviation of 1000 Bottleneck distances. $^i$ Two samples of size $n$ each one from the same object of two distinct circles. The two distinct circles have a distance of 1.5 for each point; $0.4n$ points were taken from a circle with radius $r=0.5$, and $0.6n$ were taken from a circle with radius $r=1.2$. $^j$ Two samples of size $n$ each one from different objects of two distinct circles. The first sample was taken from the previous object of two distinct circles. The second sample was taken from two distinct circles having a distance of 4.5 for each point; $0.4n$ points were taken from a circle with radius $r=1.2$, and $0.6n$ were taken from a circle with radius $r=4$.}

\normalsize
\newpage
\begin{center}
\fontsize{6.4}{1.16}\selectfont
\begin{tabular}{l|lcc|ccc|cc|cccc}

\mc{10}{c}{\bf{Table A.3.} Sampling From Two Concentric Circles} \\
\\
\\
\\
\\
\\
\\
\\
\\
\\
\\
\\
\\
\\
\\
\boldmath{$n$$^a$} && \textbf{Bottleneck$^b$}&&& \textbf{ Wasserstein$^c$} &&\textbf{ Ratio$^d$}&&&\textbf{ RST $^e$}\\
\\
\\
\\
\\
\\
\\
\\
\\
& \textbf{Range $^f$}& \textbf{IQR $^g$} & \textbf{Std $^h$}& \textbf{Range}& \textbf{IQR} & \textbf{Std}& \textbf{min}&\textbf{max}& \textbf{0}&\textbf{1}&\textbf{2}&\textbf{3}\\
\\
\\
\\
\\
\\
\\
\\
\\
\\
\\
\\
\\
\\
\mc{10}{c} {\bf Two Samples of Two Concentric Circles $^i$} \\
\\
\\
\\
\\
\\
\\
\\
\\
\\
\\
\\
\\
\\
\textbf{100} &[0.065,0.554]	&[0.107,0.188]&0.076&[0.097,0.659]&[0.192,0.280]&0.082&0.67&0.84&0.919&0.079&0.002&0\\
\\
\\
\\

\textbf{300}&[0.039,0.304]	&[0.067,0.104]&0.035&[0.087,0.286]&[0.132,0.166]&0.028&0.44&1.06&0.959&0.038&0.003&0\\
\\
\\
\\
\\

\textbf{500}&[0.032,0.236]	&[0.054,0.083]&0.026&[0.069,0.201]&[0.106,0.129]&0.018&0.46&1.17&0.971&0.029&0&0\\
\\
\\
\\

\textbf{1000}&[0.026,0.146]	&[0.040,0.061]&0.017&[0.054,0.135]&[0.074,0.089]&0.011&0.48&1.08&0.953&0.045&0.002&0\\
\\
\\
\\
\textbf{1500}&[0.018,0.119]	&[0.033,0.049]&0.014&[0.041,0.103]&[0.058,0.069]&0.008&0.44&1.15&0.962&0.031&0.007&0\\
\\
\\
\\
\textbf{2000}&[0.018,0.090]	&[0.030,0.042]&0.011&[0.034,0.082]&[0.048,0.057]&0.007&0.52&1.10&0.959&0.035&0.006&0\\
\\
\\
\\
\textbf{2500}&[0.015,0.102]	&[0.026,0.037]&0.010&[0.032,0.076]&[0.042,0.050]&0.006&0.46&1.34&0.955&0.041&0.004&0\\
\\
\\
\\
\textbf{3000}&[0.015,0.079]	&[0.024,0.034]&0.009&[0.027,0.068]&[0.038,0.045]&0.005&0.55&1.17&0.952&0.046&0.002&0\\
\\
\\
\\
\textbf{20000}&[0.006,0.029]	&[0.010,0.013]&0.003&[0.016,0.026]&[0.018,0.020]&0.002&0.37&1.10&0.920&0.066&0.014&0\\
\\
\\
\\
\textbf{25000}&[0.006,0.029]	&[0.009,0.012]&0.003&[0.015,0.024]&[0.018,0.019]&0.001&0.37&1.17&0.927&0.053&0.020&0\\
\\
\\
\\
\textbf{30000}&[0.005,0.025]&[0.008,0.011]&0.003&[0.014,0.023]&[0.017,0.018]&0.001&0.36&1.08&0.913&0.067&0.019&0.001\\
\\
\\
\\
\\
\\
\\
\\
\\
\\
\\
\\
\\
\\
\\
%
%
%
%
\mc{10}{c} {\bf Two Samples of Different Two Concentric Circles $^j$} \\
\\
\\
\\
\\
\\
\\
\\
\\
\\
\\
\\
\textbf{100} &[0.088,0.619]	&[0.165,0.281]&0.094&[0.147,0.825]&[0.289,0.432]&0.116&0.59&0.75&0.060&0.594&0.341&0.005\\
\\
\\
\\
\textbf{300} &[0.090,0.415]	&[0.140,0.204]&0.050&[0.126,0.425]&[0.214,0.273]&0.045&0.71&0.98&0.014&0.550&0.431&0.005\\
\\
\\
\\
\textbf{500} &[0.082,0.309]	&[0.128,0.185]&0.041&[0.125,0.348]&[0.170,0.211]&0.030&0.66&0.89&0.004&0.438&0.552&0.006\\
\\
\\
\\
\textbf{1000} &[0.078,0.241]&[0.125,0.169]&0.032&[0.086,0.205]&[0.119,0.143]&0.017&0.91&1.17&0.004&0.347&0.641&0.008\\
\\
\\
\\
\textbf{1500} &[0.075,0.230]	&[0.126,0.162]&0.026&[0.072,0.156]&[0.097,0.114]&0.013&1.04&1.47&0&0.361&0.629&0.010\\
\\
\\
\\
\textbf{2000} &[0.083,0.206]	&[0.126,0.157]&0.022&[0.057,0.124]&[0.082,0.096]&0.010&1.46&1.66&0&0.311&0.683&0.006\\
\\
\\
\\
\textbf{2500} &[0.081,0.203]&[0.125,0.152]&0.021&[0.057,0.114]&[0.073,0.084]&0.009&1.42&1.78&0.001&0.329&0.664&0.006\\
\\
\\
\\
\textbf{3000} &[0.082,0.189]&[0.126,0.151]&0.019&[0.050,0.096]&[0.065,0.076]&0.008&1.66&1.96&0.002&0.339&0.653&0.006\\
\\
\\
\\
\textbf{20000} &[0.108,0.151]	&[0.128,0.136]&0.007&[0.028,0.040]&[0.033,0.036]&0.002&3.82&3.73&0&0.412&0.577&0.011\\
\\
\\
\\
\textbf{25000} &[0.114,0.148]	&[0.127,0.136]&0.006&[0.027,0.037]&[0.031,0.034]&0.002&4.18&3.98&0&0.448&0.543&0.009\\
\\
\\
\\
\textbf{30000} &[0.113,0.147]&[0.128,0.135]&0.005&[0.027,0.036]&[0.030,0.033]&0.002&4.26&4.08&0.001&0.451&0.541&0.007\\

\end{tabular}
\end{center}
\footnotesize{$^a$ Sample size. $^b$ Bottleneck distance between 1000 pairs of $H_0$ persistence diagrams corresponded to 1000 paired samples. $^c$ Wasserstein distance between 1000 pairs of $H_0$ persistence diagrams corresponded to 1000 paired samples. $^d$ Ratio of the Bottleneck range to the Wasserstein range; ratio of the lowest edge ("min"), and ratio of the highest edge("max"). $^e$ Proportion of 0,1,2,3 significant differences of the RST parameters, over 1000 pairs models fitted to 1000 pairs of $H_0$ persistence diagrams. $^f$ Range of 1000 Bottleneck distances. $^g$ Interquartile range of 1000 Bottleneck distances. $^h$ Standard deviation of 1000 Bottleneck distances. $^i$ Two samples of size $n$ each one from the same object of two concentric circles; $0.4n$ points were taken from a circle with radius $r=1$, and $0.6n$ points were taken from a circle with radius $r=2$. $^j$ Two samples of size $n$ each one from the same object of two concentric circles; $0.4n$ points were taken from a circle with radius $r=2$, and $0.6n$ points were taken from a circle with radius $r=4$.}

\normalsize
\newpage
\begin{center}
\fontsize{6.4}{1.16}\selectfont
\begin{tabular}{l|lcc|ccc|cc|cccc}

\mc{10}{c}{\bf{Table A.4.} Two Distinct Circles vs. Two Concentric Circles $^a$} \\
\\
\\
\\
\\
\\
\\
\\
\\
\\
\\
\\
\\
\\
\\
\boldmath{$n$$^b$} && \textbf{Bottleneck$^c$}&&& \textbf{ Wasserstein$^d$} &&\textbf{ Ratio$^e$}&&&\textbf{ RST $^f$}\\
\\
\\
\\
\\
\\
\\
\\
\\
& \textbf{Range $^g$}& \textbf{IQR $^h$} & \textbf{Std $^i$}& \textbf{Range}& \textbf{IQR} & \textbf{Std}& \textbf{min}&\textbf{max}& \textbf{0}&\textbf{1}&\textbf{2}&\textbf{3}\\
\\
\\
\\
\\
\\
\\
\\
\\
\\
\\
\\
\\
\\
\\
\\
\\
\textbf{100} &[0.079,0.585]	&[0.131,0.215]&0.078&[0.097,0.659]&[0.192,0.280]&0.082&	0.81&0.89&0.799&	0.192&	0.009&	0\\
\\
\\
\\

\textbf{300}&[0.066,0.280]&	[0.095,0.135]&0.036&[0.087,0.286]&[0.132,0.166]&0.028&0.75&0.98&0.261&0.545&0.189&0.005\\
\\
\\
\\
\\

\textbf{500}& [0.069,0.282]&[0.092,0.118]&	0.023&[0.069,0.201]&[0.106,0.129]&0.018&1.00&1.40&0.233&0.545&0.220&0.002\\
\\
\\
\\

\textbf{1000}& [0.064,0.186]&[0.100,0.126]&0.020&[0.054,0.135]&[0.074,0.089]&0.011&1.19&1.38&	0.117&	0.605&	0.275&	0.003\\
\\
\\
\\
\textbf{1500}& [0.055,0.162]&[0.097,0.125]&0.020&[0.041,0.103]&[0.058,0.069]&0.008&1.34&1.57&	0.083&	0.639&	0.269&	0.009\\
\\
\\
\\
\textbf{2000}& [0.055,0.161]&[0.098,0.123]&0.019&[0.034,0.082]&[0.048,0.057]&0.007&	1.62&1.96&0.051&	0.629&	0.314&	0.006\\
\\
\\
\\
\textbf{2500}& [0.063,0.167]&[0.096,0.120]&0.017&[0.032,0.076]&[0.042,0.050]&0.006&1.99&2.18&	0.034&	0.573&0.383&0.010\\
\\
\\
\\
\textbf{3000}& [0.053,0.159]&[0.098,0.118]&0.016&[0.027,0.068]&[0.038,0.045]&0.005&1.96&2.35&0.034&0.573&0.382&0.011\\
\\
\\
\\
\textbf{20000}& [0.080,0.119]&[0.096,0.104]&0.006&[0.016,0.026]&[0.018,0.020]&0.002&5.15&4.57&0.005&0.497&0.493&0.005\\
\\
\\
\\
\textbf{25000}& [0.082,0.116]&[0.096,0.103]&0.005&[0.015,0.024]&[0.018,0.019]&0.001&5.36&4.74&0.005&0.485&0.506&0.004\\
\\
\\
\\
\textbf{30000}& [0.083,0.115]&[0.096,0.102]&0.005&[0.014,0.023]&[0.017,0.018]&0.001&5.78&4.91&0.008&0.446&0.539&0.007\\
\end{tabular}
\end{center}
\footnotesize{$^a$ One sample includes $n$ points from two distinct circles having a distance of 4.5 for each point; $0.4n$ points were taken from a circle with radius $r=1.2$, and $0.6n$ points were taken from a circle with radius $r=4$. The second sample includes $n$ points from two concentric circles; $0.4n$ points were taken from a circle with radius $r=1$, and $0.6n$ points were taken from a circle with radius $r=2$. $^b$ Sample size. $^c$ Bottleneck distance between 1000 pairs of $H_0$ persistence diagrams corresponded to 1000 paired samples. $^d$ Wasserstein distance between 1000 pairs of $H_0$ persistence diagrams corresponded to 1000 paired samples. $^e$ Ratio of the Bottleneck range to the Wasserstein range; ratio of the lowest edge ("min"), and ratio of the highest edge("max"). $^f$ Proportion of 0,1,2,3 significant differences of the RST parameters, over 1000 pairs models fitted to 1000 pairs of $H_0$ persistence diagrams. $^g$ Range of 1000 Bottleneck distances. $^h$ Interquartile range of  1000 Bottleneck distances. $^i$ Standard deviation of 1000 Bottleneck distances. }

\end{document}